\documentclass[letterpaper]{sig-alternate-shuang}
\usepackage{shuang}
\usepackage{cases}
\usepackage{amsfonts}
\usepackage{amsmath}
\usepackage{amssymb}
\usepackage{graphicx}
\usepackage{subfigure}
\usepackage{algorithm}
\usepackage{algorithmic}
\usepackage{epstopdf}
\usepackage{hyperref}
\renewcommand{\paragraph}[1]{%
  \smallskip
  \noindent{\bfseries #1}}

\begin{document}
\title{Collaborative Competitive filtering II:\\ Optimal Recommendation and Collaborative Games}
\numberofauthors{1} \author{
\alignauthor Shuang Hong Yang\\
        \affaddr{College of Computing,}\\
        \affaddr{Georgia Tech}\\
        \affaddr{Atlanta, GA 30318}\\
        \email{shy@gatech.edu}
}

\maketitle

\begin{abstract}
Recommender systems have emerged as a new weapon to help online
firms to realize many of their strategic goals (e.g., to improve
sales, revenue, customer experience etc.). However, many existing
techniques commonly approach these goals by seeking to recover
preference (e.g., estimating ratings) in a matrix completion
framework. This paper aims to bridge this significant gap between
the clearly-defined strategic objectives and the
not-so-well-justified proxy.

We show it is advantageous to think of a recommender system as an
analogy to a \emph{monopoly economic market} with the system as the
sole \emph{seller}, users as the \emph{buyers} and items as the
\emph{goods}. This new perspective motivates a game-theoretic
formulation for recommendation that enables us to identify the
optimal recommendation policy by explicit optimizing certain
strategic goals. In this paper, we revisit and extend our prior
work, the \emph{Collaborative-Competitive Filtering} preference
model \cite{YanLonSmo11}, towards a game-theoretic framework. The
proposed framework consists of two components. First, a conditional
preference model that characterizes how a user would respond to a
recommendation action; Second, knowing in advance how the user would
respond, how a recommender system should act (i.e., recommend)
strategically to maximize its goals. We show how objectives such as
click-through rate, sales revenue and consumption diversity can be
optimized explicitly in this framework. Experiments are conducted on
a commercial recommender system and demonstrate promising results.
\end{abstract}

\vspace{1mm} \noindent
{\bf Categories and Subject Descriptors}\\
{H.5.3} [{Information Systems}]: {\it Web-based Interaction};\\
{H.3.3} [{Information Search and Retrieval}]: {\it {Information
filtering}}

\vspace{1mm} \noindent {\bf General Terms}\\ Algorithms, Performance

\vspace{1mm} \noindent {\bf Keywords}: Recommendation optimization,
Collaborative games, Econometric model, Expected utility theory

\section{Introduction}
\label{sec:intro}
Recommender systems have become a core component for today's online
businesses. With the abilities of connecting merchant \emph{supply}
(i.e., items of various types such as retailing products, movies,
articles, ads, experts, etc.) to market \emph{demands} (i.e.,
potentially interested consumers), recommender systems are helping
online firms (e.g. Amazon, Netflix, Yahoo!) to realize many of their
hard-to-attain business goals (e.g., to boost sales, improve
revenue, enhance customer experiences)
\cite{BryHuSmi03,BryHuSim11,FleHos07,TanNet10}. Compared to an
offline market, online recommender system has the unbeatable
convenience in control, intervention, monitoring and measurement of
the market, and consequently the appealing opportunity to adjust its
operational actions (i.e., recommendation policy) to optimize
certain strategic objectives. Surprisingly, despite the fact that
many of these goals are clearly defined, they are not optimized in
today's recommender systems in a well-justified way. Instead,
research on recommendation has been focused almost exclusively on
learning preference (e.g., estimating a user's rating to a movie) in
a matrix completion formulation
\cite{SarwarWWW01,McLaughlinSIGIR04,KorenIEEE09,ChenKDD09,WeiKarLe07}.
It is rather unclear how preference learning, as a proxy,
approximates these goals, or how a strategic intervention should be
designed to achieve certain goals.

In this paper, we seek to bridge this significant gap. We show it is
advantageous to look at the \emph{user-system interactions} and
think of a recommender system as an analogy to a \emph{monopoly
economic market} (i.e., system as the sole seller, users as buyers
and items as goods)\footnote{Hereafter, we will use interchangeably
``system" and ``seller", ``user" and ``buyer", ``item" and
``good".}, rather than \emph{user-item interactions} as in the
conventional \emph{matrix completion} formulation. This new
perspective motivates a novel game-theoretic formulation, upon which
recommendation policy can be optimized strategically with respect to
business objectives such as click-through rate, sales revenue and
consumption diversity.

\subsection{User-System Interactions}
Recommender systems are commonly designed by analyzing the dyadic
\emph{user-item interactions} as can be recorded by a matrix, for
example, users assigning ratings to movies. Research has thus been
focused exclusively on estimating preference or equivalently
completing the matrix of ``who-like-what"
\cite{BalSho97,SarwarWWW01,McLaughlinSIGIR04,WeiKarLe07,AiroldiNIPS08,AgaChe09,KorenIEEE09,ChenKDD09}.
This \emph{matrix-completion} formulation of recommendation has been
extensively investigated and become especially popular thanks to the
Netflix Prize Competition. Nonetheless, as we show in this paper,
the formulation of recommendation as user-item interaction or matrix
completion is inherently flawed --- recommendation is not solely
about \emph{what you know} (i.e., knowledge about the user), but
more importantly about \emph{how you act} (i.e., how to recommend
items to serve the user or persuade the user to consume). Instead,
it is advantageous to think of recommendation as an interaction
between the system and the users and formulate it as an
interdependent decision-making process (aka \emph{games})
\cite{LeySho08}.

In a typical interaction, the system \emph{acts} by providing a set
of personalized recommendations, and user \emph{reacts} by making
choices, i.e., by choosing to consume some of the recommended items
(e.g., click a link, rent a movie, view a News article, purchase a
product). This process in many aspects resembles what happens in a
monopoly market where the recommender system, as the sole seller,
has absolute market power to manipulate the market, yet the utility
it receives depends on the reaction of the buyers (i.e., users),
e.g., the success of an advertising system is directly related to
how users react (i.e., whether they click the ads or not). Clearly,
the action of the seller and the reaction of the buyer are
interdependent -- the two players (i.e., seller and buyer) each has
its own objective (i.e., utility) to achieve, yet how and to what
extend they can achieve their own objectives depends also on the
decision of the other player. Conventional matrix completion
formulation for recommendation is inherently flawed as it is
incapable to capture such interdependent decision making
interactions. As a results, although many business objectives in
e-commerce are clearly defined, how a recommender can be designed to
optimize these goals hasn't yet been explored.

\subsection{Recommendation as Collaborative Games}
In this paper, we present a game-theoretic formulation for
recommendation, where the user-system interactions are modeled as a
collection of coupled games with each game played between the seller
and one buyer (i.e., between the system and one user). For the sake
of statistical inference, it is nonetheless important \emph{not} to
model these games as mutually independent.  We therefore bring
forward the notion of ``collaborative games" that \emph{similar
games are expected to yield similar outcomes}, which enables us to
pool the sparse data across games to obtain reliable statistical
estimation.

We extend our prior work on ``Collaborative-Competitive Filtering"
(CCF) preference model \cite{YanLonSmo11} towards a game-theoretic
framework. The framework consists of two components: (1) a
conditional model $p_u(R|A)$ that characterizes the reaction $R$ of
a buyer $u$ in the context of any given action $A$ of the seller;
this model enables us to predict in advance what the outcome of a
game would be (e.g., how the buyer would respond); and (2) given
$p_u(R|A)$ for every buyer $u$, a formulation for optimizing the
seller's action policy $A$ w.r.t. a predefined payoff (e.g., a
strategic goal).

To effectively model $p_u(R|A)$, we revisit and extend the CCF
preference model \cite{YanLonSmo11} that integrates latent factor
models in collaborative filtering with choice models in
econometrics. By using latent factor based utility parametrization,
the model encodes the ``collaboration effects" among games
\cite{Byr71,McpLovCoo01} to advocate the notion of ``collaborative
games". As the policy spaces are prohibitively large yet the
observations are extremely sparse, this formulation is essential for
reliable statistical inference because it enables the sparse data to
be pooled across games. It also remarkably reduces the parametric
complexity of $p_u(R|A)$ significantly from a prohibitive high-order
polynomial scale down to a linear scale.

The knowledge about users' reaction behavior, as characterized by
$p_u(R|A)$, enables us to predict ``future" (i.e., user's reaction)
with uncertainty and further to optimize the action (i.e., the
recommendation policy) of the recommender system strategically
\cite{LeySho08}. For any input action $A$, the possible outcomes of
the games occur with probabilities defined upon $p_u(R|A)$. Given a
payoff (i.e., a function of the outcome) that is von
Neumann-Morgenstern rational, the expected utility theory asserts
that the best action is the one that maximizes the expected payoff
\cite{VNM}. We show how business objectives such as click-trough
rate, sales revenue and consumption diversity can be formulated
explicitly as expected utilities and used in turn to optimize a
recommender system's action policy.

We also show that the CCF model is sequentially rational and thus
approximates the \emph{perfect Nash equilibrium} \cite{LeySho08}.
Experiments on a real-world commercial system demonstrate that the
proposed CCF model not only outperforms CF models in both offline
and online tests but is also highly effective in achieving
satisfactory strategic goals.

\paragraph{Outline:}
The rest of the paper is structured as follows. We first briefly
review the current matrix completion formulation and collaborative
filtering in Section~\ref{sec:cf}. We then present our new
game-theoretic formulation in Section~\ref{sec:usi} and the CCF
model in Section~\ref{sec:base}. Experiments are presented in
Section~\ref{sec:exp}, followed by summary and conclusion in
Section~\ref{sec:sum}.

\section{User-Item Interactions and\\ Collaborative Filtering}
\label{sec:cf}

Many existing approaches generally think of recommendation as
\emph{user-item interactions} and therefore aim to recover/estimate
the preference of each individual user to the items. Given a set of
$N$ users
\[u \in \mathcal{U}:=\{1, 2, \ldots, N\}\]
and a set of $M$ items
\[i \in \mathcal{I}:=\{1,2,\ldots, M\},\]
this
is naturally formulated as a matrix completion problem, where we are
given observations of dyadic responses $\{(u,i,y_{ui})\}$ with each
$y_{ui}$ being an observed response indicating user's preference
(e.g.\ user's rating to an item, or indication of whether user $u$
likes item $i$), the goal is to complete the whole mapping:
\begin{align*}
  (u,i) \rightarrow y_{ui}\text{ where } u\in \mathcal{U}, i\in
  \mathcal{I}
\end{align*}
which constitutes a large matrix $Y \in
\mathcal{Y}^{\vert\mathcal{U}\vert\times \vert\mathcal{I}\vert}$.
Assume each item can be consumed multiple times, recommendations are
usually done by a simple preference-based ranking according to $Y$,
(i.e., recommending the items with highest $y_{ui}$ scores to user
$u$). This formulation include both of the two major categories of
approaches to recommendation, i.e., content-based filtering
\cite{BalSho97,ChenKDD09} and collaborative filtering
\cite{SarwarWWW01,McLaughlinSIGIR04,AgaChe09,KorenIEEE09}, among
which we briefly review the latter.

It is worth noting that the observed responses are often extremely
sparse in realistic systems, i.e., while we might have millions of
users and items, only a tiny proportion (considerably less than 1\%)
of the entries of the matrix $Y$ are observable. This ``data
sparseness" issue has been widely recognized as one of the key
challenges of recommender system
\cite{McLaughlinSIGIR04,AgaChe09,KorenIEEE09}. To this end,
\emph{collaborative filtering} (CF) explores the notion of
``collaboration effects", i.e., similar users have similar
preferences to similar items. By encoding collaboration, CF pools
the sparse observations in such a way that for predicting
$\hat{y}(u,i)$ it also borrows observations from other (similar)
users/items. Generally speaking, existing CF methods fall into
either of the following two categories.

\paragraph{Neighborhood models.}
A popular class of approaches to CF is based on propagating the
observations of responses among items or users that are considered
as neighbors. The model first defines a similarity measure
between items / users. Then, an unseen response between user $u$ and
item $i$ is approximated based on the responses of neighboring users
or items \cite{SarwarWWW01,McLaughlinSIGIR04}, for example, by
simply averaging the neighboring responses with similarities as
weights.

\paragraph{Latent factor models.}
This class of methods learns predictive latent factors to estimate
the missing dyadic responses. The basic idea is to associate latent
factors\footnote{We assume each latent factor $\phi$ contains a
constant component so as to absorb user/item-specific offset into
the latent factor $\phi$ and $\psi$.}, $\phi_u \in\mathbb{R}^k$ for
each user $u$ and $\psi_i\in\mathbb{R}^k$ for each item $i$, and
assume a multiplicative model for the dyadic response,
\begin{align}
  p(y_{ui}|u,i) = p(y_{ui} | \phi_u^\top \psi_i; \Theta),\notag
\end{align}
where $\Theta$ denotes the set of hyper-parameters. This way the
factors could explain past responses and in turn make prediction for
future ones. This model implicitly encodes the Aldous-Hoover theorem
\cite{Kallenberg05} for exchangeable matrices -- $y_{ui}$ are
independent of each other given $\phi_u$ and $\psi_i$. In essence,
it amounts to a low-rank approximation of the matrix $Y$ that
naturally embeds both users and items into a vector space in which
the inner-products directly reflect the semantic relatedness.

To design a concrete model
\cite{AiroldiNIPS08,AgaChe09,KorenIEEE09,MillerNIPS09,SinghECML08},
one needs to specify a distribution for the dependence. Afterwards,
the model boils down to an optimization problem. For example two
commonly-used formulations are:
\begin{description*}
\item[- $\ell_2$ regression] The most popular learning formulation is to
  minimize the $\ell_2$ loss within an empirical risk minimization
  framework \cite{KorenIEEE09}:
  \begin{align*}
\min_{\phi,\psi} \sum_{(u,i)\in \mathcal{O}} (y_{ui} -
\phi_u^\top\psi_i)^2 +
\lambda_{\mathcal{U}}\sum_{u\in\mathcal{U}}\vert\vert\phi_u\vert\vert^2
+
\lambda_{\mathcal{I}}\sum_{i\in\mathcal{I}}\vert\vert\psi_i\vert\vert^2,
\end{align*}
  where $\mathcal{O}$ denotes the set of ($u,i$) dyads for which the
  responses $y_{ui}$ are observed, $\lambda_{\mathcal{U}}$ and
  $\lambda_{\mathcal{I}}$ are regularization weights.
\smallskip\item[- Logistic] Another popular formulation
  \cite{MillerNIPS09,AgaChe09} is to use logistic regression by
  optimizing the cross-entropy:
\end{description*}
\vspace{-10pt}\begin{align*}
    \min_{\phi,\psi} & \sum_{(u,i)\in \mathcal{O}}
    \log \sbr{1 + \exp(-\phi_u^\top \psi_i)} + \lambda_{\mathcal{U}}\sum_{u\in\mathcal{U}}\vert\vert\phi_u\vert\vert^2
    + \lambda_{\mathcal{I}}\sum_{i\in\mathcal{I}}\vert\vert\psi_i\vert\vert^2
  \end{align*}

\section{User-System Interaction as\\ Collaborative Games}
\label{sec:usi}
Based on the perspective of u\emph{ser-item interactions}, the
matrix completion formulation for recommendation has led to numerous
algorithms which excel at a number of data sets, including the
prize-winning work of \cite{KorenIEEE09} and many other successful
collaborative filtering algorithms
\cite{SarwarWWW01,McLaughlinSIGIR04,SalMni08,AgaChe09,KorenIEEE09,WeiKarLe07,LiuYan08}.
However, as we discussed, this formulation is inherently flawed;
instead, it is advantageous to model the \emph{user-system
interactions} so as to capture the interdependent decision-making
process between the system and the users. This motivates a novel
game-theoretic formulation for recommendation and opens up a
promising direction that enable us to optimize recommendation policy
strategically in respect of important business objectives, which
cannot be achieved otherwise with the conventional matrix completion
formulation.

\begin{table}
    \centering
    {
      \begin{tabular}{c|c|c}
        \hline User   & System Action               & User Reaction \\
        \hline $u_1$  & [$i_1,i_2,i_3,i_5$]     & $i_2$ \\
        \hline $u_2$  & [$i_2,i_3,i_4,i_5$]     & $\emptyset$ \\
        \hline $u_3$  & [$i_1,i_3,i_5,i_6$]     & $i_5$ \\
        \hline $u_4$  & [$i_2,i_3,i_4,i_6$]     & $i_3$ \\
        \hline $u_5$  & [$i_1,i_3,i_4,i_5$]     & $i_4$ \\
        \hline $u_6$  & [$i_1,i_4,i_5,i_6$]     & $i_6$ \\
        \hline
      \end{tabular}}
  \caption{An example trace of user-system interactions in recommendation.
    \label{tab:example}}
\end{table}

Consider a typical scenario of user-system interaction in a
recommender system: we have $N$ users $u \in \mathcal{U}:=\{1, 2,
\ldots, N\}$ and $M$ items $i \in \mathcal{I}:=\{1,2,\ldots, M\}$;
when a user $u$ visits the site, the system recommends a set of
items $A = \{i_1,\ldots, i_l\}$ and $u$ in turn chooses a (possibly
empty) subset $R\subseteq A$ for consumption (e.g.\ buys some of the
recommended products). From now on, we refer to $A$ as
\emph{action}, and $R$ as \emph{reaction}. For simplicity, we assume
each action is fixed-size with a given length, $|A|=l$, and that
each reaction is either empty or contains exactly one choice,
$|R|=1$ or 0. Therefore, we have $A\in \mathcal{A}=\mathcal{I}^l$
and $R\in
\mathcal{R}\subset\tilde{\mathcal{I}}=\mathcal{I}\cup\{\emptyset\}$.
Table~1 shows an example trace of such interactions.

The behavior of the recommender system and that of the users are
interdependent. On the one hand, since people make different
decisions when facing different contexts, a user's decision $R$
depends crucially on the action of the system, $A$, (i.e., what was
provided to him). For instance, an item $i$ would not have been
chosen by $u$ if it were not presented to him at the first place;
likewise, user $u$ could choose another item if the context $A$
changes such that a better item were recommended to him. On the
other hand, how a recommender system acts also depend on user's
behavior (i.e., response), because the success of recommendation
(i.e., in terms of click-through, revenue, etc.) is defined directly
on how users react to it (e.g., purchase a product, click an ad,
rent a movie). It is therefore nature to formulate recommendation
based on game theory, as analogy to a monopoly market where the
recommender as the sole seller, a user as a buyer and the items as
the goods.

Formally, the user-system interactions in a recommender system can
be formulated as a set of $N$ non-cooperative games
$\mathcal{G}=\{G_n=(P_n,\mathcal{Z}_n,U_n),\text{ }
n=1,2,\ldots,N\}$. For each game $G_n$, the player set
$P_n=\{S,u_n\}$ consists of two players, i.e., the system (i.e.,
seller) $S$ and a user (i.e., buyer) $u_n$; the policy space
$\mathcal{Z}_n=\mathcal{A}\times\mathcal{R}\subset\mathcal{I}^l\times\tilde{\mathcal{I}}$
is the set of all possible action-reaction pairs $Z_n=(A_n,R_n)$,
where $Z$ is called an outcome and $\mathcal{Z}$ the outcome space;
and the utility (i.e., payoff) function $U_n=\{U_S(Z_n), U_u(Z_n)\}$
consists of the system's payoff $U_S$ and the user's payoff $U_u$.
At an interaction $t$, a user $u_t$ visits the system and the game
$G_{u_t}$ is played with outcome $Z_t=(A_t,R_t)$ and utility output
$U(A_t,R_t)$. Since the users' behavior is not in our control, our
goal in designing a recommender system is to generate a system
action (recommendations) $A_{\tilde{t}}$ for an incoming visit
$\tilde{t}$ of user $u_{\tilde{t}}$ so as to maximize the system's
payoff $U_s(Z_{\tilde{t}})$.

It is important to emphasize that the games in $\mathcal{G}$ should
\emph{not} be modeled as independent games. Particularly, since the
outcome space can be very large, yet observations are typically
sparse, it is practically important to still be able to leverage the
\emph{collaboration effect} such that similar games are expected to
yield similar outcomes. This way it enables us to pool the sparse
evidences across different but similar games and in turn obtain
reliable statistical inference. For this reason, we term the
formulation ``\emph{collaborative games}" with a slight abuse of
terminology.

This game-theoretic formulation provides a novel perspective for
recommendation. Particularly, since the strategies of the buyer and
the seller are interdependent, to optimize the seller's action, we
have to (1) for each candidate action $A$, predict the buyer's
reaction $R$ in advance; and then (2) find the best action $A$ by
maximizing the achievable payoff $U_s(A,R)$.

\section{Collaborative competitive\\ filtering}
\label{sec:base}
Our recent work \cite{YanLonSmo11} established the first principled
model for learning preference from user-system interactions in
recommendation system. Unlike conventional preference learning
models which are trained on the who-like-what matrix, our CCF
preference model is trained on user-system interactions where the
system action $A$ is used as a context in which a user's reaction
(e.g., ``like") $R$ is made; in other word, CCF model doesn't only
capture who-like-what, but it also considers what are the options
available to the user when the ``like" decision is made. As
demonstrated in our experiments \cite{YanLonSmo11} and many other
successful applications (e.g., online test on Yahoo! and Netflix),
the CCF preference model significantly improves recommendation
performance on a variety of data sets. However, like many existing
recommendation algorithms, our prior CCF model is still within the
conventional matrix completion framework. To be precise, all these
models only care about, and are only capable to model, the behavior
of the user (i.e., what a user likes). These techniques are lacking
as they largely ignore the interdependent or game-theoretic nature
of the user-system interactions in recommendation, and consequently,
none of them is able to optimize the recommendation policy
explicitly in respect of a prescribed objective (although many
strategic objectives for a recommender system are clearly defined).

In this paper, we extend our prior work and present a game-theoretic
framework for recommendation. We would like to keep the name
``Collaborative-Competitive Filtering" or CCF since the preference
model we established in our prior work is revised and used as one
essential component of this framework. The CCF framework consists of
two components: (1) a model $P_u(R|A)$ that predicts in advance
(with uncertainty) a buyer's reaction $R$ to a given action $A$; and
(2) a formulation for finding the best action strategy (i.e.,
recommendation policy) for the seller.

\subsection{Conditional User Reaction Modeling}
\label{sec:choice}
The first part of the framework is to predict a buyer's choice $R$
in the context of any given action $A$ of the seller's. In a
decision environment with imperfect information, this means to
quantify the conditional distribution $p_u(R\vert A)$. The full
parametrized version of this distribution requires $O(NM^{l+1})$
free parameters, statistical estimation of which is practically
prohibitive since the observations are typically available only at a
scale far less than $O(NM)$ (e.g., in matrix completion, usually
less than $1$\% entries are observed). In this section, we revisit
and extend our CCF preference model \cite{YanLonSmo11} by presenting
a conditional reaction model with complexity $O(N+M)$.

\subsubsection{Behavioral Axioms of Choice Process}
We first present an axiomatic view of the choice process. We assume
a good (i.e., item) $i$ has a potential utility $r_{ui}$ to a buyer
$u$. Moreover, we assume a buyer $u$ is a rational decision maker:
he knows that his choice of a good $i$ will be at the expense of
other available alternatives $i^\prime\in A$, therefore he compares
among all the alternatives before making his choice. In other words,
for each decision, $u$ considers both \textsf{revenue} and
\textsf{opportunity cost}, and decides which good to buy based on
the potential \textsf{profit} of each good in $A$. Specifically, the
opportunity cost $c_{ui}$ is the potential loss of $u$ from buying a
good $i$ that excludes him to buy other alternatives: $c_{ui} =
\max\{r_{ui^\prime}: i^\prime\in A\setminus i\}$; the profit
$\pi_{ui} = r_{ui} - c_{ui}$ is the net gain of an decision. Based
on the rational decision theory \cite{Luc59}, we have the following
axiom about the buyer's choice reaction.

\smallskip\noindent{\textsc{{{Axiom 1} [Local optimality of choice]}}}: \textit{A
rational decision is a decision maximizing the profit: $i^*$ =
$\arg\max_{i\in A} \pi_{ui}$.}\smallskip

This axiom implies a \emph{local competitive effect}: the buyer $u$
turns to chooses the good that is locally the best in the context of
the available alternatives in $A_t$. Unfortunately, the axiom
restricts the utility function only up to an arbitrary
order-preserving transformation (e.g.\ a monotonically increasing
function), and hence cannot yield a unique solution \cite{Man75}.
Another issue is that it is deterministic, less useful since we
don't have perfect information about how users react. To this end,
we draw an stochastic counterpart of this axiom from the random
utility theory \cite{Luc59,McF73}:

\smallskip\noindent{\textsc{{{Axiom 2} [Independence of Irrelevant Alternatives]}}}: \textit{For any given context
set $A$, the relative odds of a user $u$'s selecting an item $i\in
A$ over another item $j\in A$ should be independent of the presence
or absence of any irrelevant items, i.e.,}
\begin{align}
&\frac{p_u(i\vert \{i,j\})}{p_u(j\vert \{i,j\})} = \frac{p_u(i\vert
A)}{p_u(j\vert A)}
\end{align}

Note that this axiom brings the parametric complexity of $p_u(R\vert
A)$ significantly down from $O(NM^{l+1})$ to $O(NM^2)$.

\subsubsection{User Utility Parametrization}
In the spirit of the random utility theory \cite{Luc59,McF73}, we
decompose the buyer's utility function into two parts, i.e.,
$U_{u}(i) = r_{ui} + e_{ui}$, where: (1) $r_{ui}$ is a deterministic
component characterizing the intrinsic interest of the buyer $u$ to
the good $i$; (2) the second part $e_{ui}$ is a stochastic
unobserved error term reflecting the uncertainty, richness and
complexity of the choice process. Under very mild conditions, it has
been shown that the error terms $e_{ui}$ are independently and
identically distributed with the Weibull (extreme point)
distribution \cite{Gumbel54}:
\begin{align}
P(e_{ui}\leqslant\epsilon) = e^{-e^{-\epsilon}}.
\end{align}
Furthermore, to encode the collaborative effect such that the
observed evidences could be pooled across similar games, we
parametrize the deterministic utilities, $r_{ui}$, with the
multiplicative latent factor model \cite{KorenIEEE09,AgaChe09}:
\begin{align}
&r_{ui} = \phi_u^\top\psi_i
\end{align}
where $\phi_u\in\mathbb{R}^k$ and $\psi_i\in\mathbb{R}^k$ are
low-rank latent profiles for user $u$ and item $i$ respectively,
just as in the collaborative filtering models we described in
Section~\ref{sec:cf}.
\subsubsection{The Multinomial Logit Factor Model}
The behavioral axiom and the low-rank parametrization together lead
to the following theorem.

\smallskip\noindent{\textsc{{{Theorem  1}:}}} \textit{Suppose the utility
function $U_u(i)=r_{ui}+\epsilon_{ui}$, where $\epsilon$ are i.i.d.
Weibull variables, then the distribution of selecting one item that
satisfies Axiom 2 is given by $p_u(i\vert A)
  ={e^{r_{ui}}}/{\sum_{j\in A} e^{r_{uj}}}$ for any $i\in A$}.

\smallskip\noindent{\it Proof.} c.f. \cite {McF73}
.\hfill$\Box$

\medskip

The above model is well-known as the \emph{multinomial logit model},
which has been extensively used for modeling conventional offline
consumer choice behavior (e.g., choose of occupation, brand,
housing) in econometrics \cite{McF73,Man75}, sociometrics
\cite{Luc59} and marketing science \cite{GenRec79,GuaLit08}. We
adapt it for modeling online game-theoretic interactions in
recommender systems. In contrast to the traditional choice models,
where the deterministic part of the utility $r_{ui}$ is a linear
mapping $w^\top x_{ui}$ of observed features $x_{ui}$ (i.e.,
measured user and item features), here we employ the multiplicative
latent factor parametrization. The formulation proposed hereby
seamlessly integrate two distinct methodologies --- choice models in
econometrics and factorization models in collaborative filtering.
This integration is significant because it enables us to model the
seller-buyer games \emph{collaboratively}, rather than
\emph{independently} as in conventional choice models. That is, it
enables us to pool data across games such that the interactions
engaging similar users, similar actions and similar reactions are
dealt with in a similar way.

Moreover, in conventional choice models, it is assumed that in each
interaction $t$, the buyer will take at least one item $i^*\in A_t$.
This assumption is, however, not true in our case since user's visit
to a recommender system does not always yields a response. For
example, users frequently visit online e-commerce website without
making any purchase, or browse a news portal without clicking on any
ad. Actually, such nonresponded visits may account for a vast
majority of the traffics that an recommender system receives. More
interestingly, different users may have different propensities for
giving a response. It is important to reflect this in the model as
well. To this end, we add a scalar latent factor, $\theta_u$, for
each user $u$ to capture the \emph{response propensity} of the buyer
$u$. At an interaction $t$, we assume buyer $u_t$ makes an effective
purchase only if he feels that the overall quality of the offered
goods $A_t$ are good enough. In other words, there is a certain
reserve utility that needs to be exceeded for a user to respond. In
keeping with the multinomial logit model and the latent factor
parametrization, we have the following model
\begin{align}
  \label{eq:softmax_ext}
  p_u(R=i| A)
  &=\frac{\exp(\phi_u^\top\psi_{i})}{\exp(\theta_u) + \sum_{j\in
  A}\exp(\phi_u^\top\psi_j)}, \forall~i\in A;\\
  p_u(R=\emptyset| A)
  &=\frac{\exp(\theta_u)}{\exp(\theta_u) + \sum_{j\in
  A}\exp(\phi_u^\top\psi_j)}\text{ otherwise}
\end{align}
which we refer to as \emph{multinomial logit factor} or MLF model.
Note that this new formulation reduces the parametric complexity of
$p_u(R|A)$ significantly to linear scale, i.e., $O(k(N+M)+N)\approx
O(N+M)$, where $k$ is the dimensionality of the latent factor
$\phi\in\mathbb{R}^k$ and $\psi\in\mathbb{R}^k$, which is generally
a small number (usually up to a few hundreds).

\subsubsection{Position Bias}
An important factor that was overlooked by the MLF model yet is
important in practice is the \emph{position bias}. In particular,
the choice of a buyer depends not only on the utilities of the
available alternatives but also on how they are placed (i.e., the
positions), e.g., users usually pay attentions only to a few
top-ranked goods and totally disregard the others. Such position
bias is evident in many online decision making scenarios, e.g., Web
search, recommendation, advertising. We extend the MLF model by
adding a set of position-specific latent factors $\{\beta_p \in
\mathbb{R}^k, p =1,\ldots, l\}$ via:
\begin{align}
  p_u(R=i| A)
  &=\frac{\exp(\langle\phi_u,\psi_{i},\beta_{p(i)}\rangle)}{\exp(\theta_u) + \sum_{j\in
  A}\exp(\langle\phi_u,\psi_j,\beta_{p(j)}\rangle)},\label{eq:pb}
\end{align}
where $p(i)$ denotes the position of item $i$,
$\langle\phi,\psi,\beta\rangle =
1^\top(\phi\circ\psi\circ\beta)=\sum_{i=1}^k \phi[i]\psi[i]\beta[i]$
is a three-way inner product, $\circ$ denotes Hadamard (aka
element-wise) product.

\subsubsection{Conditional Maximum Likelihood Estimation}
Given a collection of training interactions $\{(u_t,A_t, R_t)\}$,
the latent factors, $\phi$ and $\psi$, can be estimated using
penalized conditional maximum likelihood estimation via
\begin{align}
  \label{eq:softmax}
& \min_{\phi,\psi,\theta, \beta}:
 \sum_{t} \{\log [e^{\theta_{u_t}}+\sum_{j\in A_t}e^{\langle\phi_{u_t}\psi_j\beta_{p_j}\rangle}] - (1-\delta_{\emptyset,t})\langle\phi_{u_t}
  \psi_{i^*}\beta_{p_{i^*}}\rangle \nonumber\\ &-
  \delta_{\emptyset,t}\theta_{u_t} \}
  + \lambda_{\mathcal{U}}\sum_{u\in\mathcal{U}}\vert\vert\phi_u\vert\vert^2
  + \lambda_{\mathcal{I}}\sum_{i\in\mathcal{I}}\vert\vert\psi_i\vert\vert^2
  + \lambda_{P}\sum_{p=1}^l\vert\vert\beta_p\vert\vert^2.\notag
\end{align}
where $\delta_{\emptyset,t} = 1$ if $R_t=\emptyset$, or 0 otherwise.

\subsubsection{Distributed Stochastic Optimization}
Due to the use of bilinear multiplications, although the conditional
likelihood is convex w.r.t. $r_{ui}$ as each of the objective terms
is strongly concave, it is nontheless nonconvex w.r.t. the latent
factors $\phi$ and $\psi$. Moreover since the interactions evolve
over time, it is desirable to have algorithms that are sufficiently
efficient and preferably capable to update dynamically so as to
reflect upcoming data streams, therefore excluding offline learning
algorithms such as classical SVD-based factorization algorithms
\cite{KorenIEEE09} or spectral eigenvalue decomposition methods
\cite{LiuYan08}. Here, we use a distributed stochastic gradient
variant based on the Hadoop MapReduce framework. The infrastructure
is analogous to what was proposed in \cite{ZinWeiSmo10}. The basic
module is a stochastic gradient descent algorithm, which loops over
all the observations and updates the parameters by moving in the
direction defined by negative gradient. For example, for a given
responded session $(u,A,i^*)$, we can carry out the following to
update the latent factors on each machine separately:
\begin{itemize*}
\item For $u$ do:
\[\phi_u \leftarrow \phi_u -
  \eta \sbr{\sum\limits_{i \in A} l'(u,i)\times \psi_i\circ\beta_{p_i}
    + \lambda_{\mathcal{U}}\phi_u}.\]
\item For each $i\in A$ do:
\[\psi_i \leftarrow \psi_i -
  \eta \sbr{l'(u,i)\times\phi_u\circ\beta_{p_i} +
  \lambda_{\mathcal{I}}\psi_i}.\]
\item For each $p\in \{1,\ldots,l\}$ do:
\[\beta_p \leftarrow \beta_p -
  \eta \sbr{l'(u,i_p)\times \psi_{i_p}\circ\phi_u
    + \lambda_{P}\beta_p}.\]
\end{itemize*}
where $\eta$ is the learning rate\footnote{We carry out an annealing
procedure to discount $\eta$ by a constant factor after each
iteration, as suggested by \cite{KorenKDD08}.}. The gradient is
given by:
\begin{align}\label{eq:grad}
l'(u,i)&=\frac{\exp(\langle\phi_u\psi_i\beta_{p_i}\rangle)}{\exp(\theta_u)+\sum_{j\in
A}\exp(\langle\phi_u\psi_j\beta_{p_j}\rangle)} - \delta_{i,i^*}.
\end{align}

\subsection{Strategic System Action Optimization}
\label{sec:actionopt}
The distribution $p_u(R|A)$ characterizes (in probability) how a
buyer would react to a given action. Knowing this enables us to
optimize the seller's action strategy (i.e., recommendation policy)
by maximizing its utility (payoff) $U_S$ \cite{LeySho08}. In this
section, we show that this can be formulated based on von
Neumann-Morgenstern's \emph{expected utility theory}. We then
specify the formulation in terms of three example payoff objectives,
i.e., click-through rate, sales revenue and consumption diversity.

\subsubsection{Expected Utility Maximization}
Because of the uncertainty/risk inherent in the game, it is nature
to formulate action optimization as \emph{decision making under
uncertainty}. Consider a given game $G_u$ between the seller $S$ and
a specific buyer $u$, the action space is the set of all possible
combinations of $l$ goods, $\mathcal{A}= \mathcal{I}^L$. An action
$A\in\mathcal{A}$ yields an outcome $Z=(A,R)\in \mathcal{Z} =
\mathcal{A}\times \mathcal{R}$ with probability distribution $p(Z)$
(\emph{aka} lottery), where the reaction space
$\mathcal{R}=A\cup\{\emptyset\}$. Because our knowledge about the
environment is imperfect, we would rather adopt a probabilistic
action strategy such that actions for $G_u$ are sampled according to
a distribution $p_u(A)$ (defined over the action space $\mathcal{A}$
, any $A\in \mathcal{A}$ is taken with probability $p_u(A)$), then
we have $p_u(Z)=p_u(A)p_u(R|A)$, where we specify the dependence on
the user with a subscript to emphasize the fact that the action is
customized for each user.

A utility function $U_S(Z)$ is a mapping
$U_S:\mathcal{Z}\rightarrow\mathbb{R}$, which defines a preference
relation $\succcurlyeq$ over the outcome space $\mathcal{Z}$ such
that $Z\succcurlyeq Z'$ if and only if $U_S(Z)\geqslant U_S(Z')$.
Without loss of generality, we assume $\succcurlyeq$ is von
Neumann-Morgenstern rational, i.e., it satisfies the four axioms:
completeness, transitivity, independence and continuity. The von
Neumann-Morgenstern (vNM) theorem defines the best outcome of a
decision in an environment under uncertainty as follows\cite{VNM}.

\smallskip\noindent{\textsc{{{Theorem  2} [Expected Utility]:}}} \textit{
Suppose $\succcurlyeq$ is a preference defined by an utility
function $U_S$ that satisfies the 4 axioms, for any two
distributions (lotteries) $p(Z)$ and $q(Z)$, we have: $p\succcurlyeq
q$ if and only if $\mathbb{E}_p(U_S)\geqslant \mathbb{E}_q(U_S)$.}

\smallskip\noindent{\it Proof.} c.f. \cite{VNM}.\hfill$\Box$

\medskip

Based on the vNM theorem, the optimal action strategy $p_u(A)$,
given $p_u(R|A)$, can be achieved by the following linear
optimization:
\begin{align}\label{eq:actopt}
&\max_{p_u(A)} \sum_{A\in\mathcal{A}}p_u(A)\sum_{R\in\mathcal{R}}p_u(R|A)U_S(A,R)\\
s.t.:& \sum_{A\in\mathcal{A}} p_u(A) = 1,\text{ and }
p_u(A)\geqslant0.\notag
\end{align}
A simplex solution for the above is given simply by:
\begin{align*}
&p_u(A)=\delta_{A,A_u^*}\text{ where }
A_u^*=\arg\max_A\sum_{R\in\mathcal{R}}p_u(R|A)U_S(A,R).
\end{align*}
In practice, it is usually favorable, (e.g., for risk-robustness
reasons) to choose a less sparse distribution (i.e., a portfolio
\cite{Mar70}) rather than the singular distribution as defined by a
simplex solution, the discussion of which is, however, beyond the
scope of this work.

\subsubsection{Action Strategy Parametrization}
\label{sec:actpara}
Although the simplex solution looks simple, exhaustive search
throughout the outcome space is still something practically
prohibitive as there are $O(NM^{l+1})$ extreme points. To this end,
we propose to parameterize the action distribution in terms of a
small set of parameters $\Theta$, e.g., to assume action $A$ is
sampled from a parametric distribution $p_u(A;\Theta)$. In this way,
we can search $\mathcal{A}$ efficiently by optimizing $\Theta$
instead. As a preliminary study, here we devise a simple
parametrization by randomizing a utility-based ranking scheme with a
scalar parameter $\alpha$. Particularly, for any given user $u$,
assume the top-ranked $l$ items (i.e., items with highest payoffs)
are denoted $\{i^*_1,\ldots,i^*_l\}$, we generate the action $A$ as
follows:

\smallskip\noindent%
\frame{\parbox{\columnwidth}{\begin{tabular}{l}
$\bullet$ $A = \emptyset$.\\
$\bullet$ For $j$ from $1$ to $l$ do:\\
\qquad- With probability $(1-\alpha)$ add $i^*_j$
to $A$\\
\qquad- With probability $\alpha$ add an random item to $A$\\
\end{tabular}}}

\smallskip\noindent%
This way, action optimization in Eq(\ref{eq:actopt}) become a
one-dimensional optimization, to which the solution can be obtained
efficiently, e.g., via golden-section search.

A more flexible parametrization is to factorize $p(A)$ sequentially
$p(A)=p(i_1)p(i_2|i_1)\ldots p(i_l|i_1,i_2,\ldots,i_{l-1})$, with a
few simplifications, we can search the action space by dynamic
programming. We leave this for future research.
\subsubsection{Strategic Payoff Specification}
So far, our discussion of action optimization is in terms of an
abstract payoff function $U_S$. We now specify our formulation with
three concrete strategic objectives.

\paragraph{Payoff \#1: Click-Through Rate (CTR).} Click-through rate
or CTR is the ratio of responses (i.e., $R_t\ne\emptyset$) out of
all the interactions. CTR is the most important measure of success
for many real-world recommender systems because it crucially
determines so many important factors ranging from traffic, revenue
to user base. For example, it corresponds to the advertisement click
rate in Google, the movie rental rate in Netflix, the order
placement rate in Amazon, and the rate of friend connection in
Facebook Friend-Finder. CTR can be formulated in the CCF framework
as follows:
\begin{align}
CTR &= \mathbb{E}_u[\mathbb{E}_A[p_u(R\ne\emptyset|A)]]\\
&=\sum_{u\in\mathcal{U}}f_u\sum_{A\in\mathcal{A}}p_u(A)p_u(R\ne\emptyset|A)\notag
\end{align}
where $f_u$ is a measure of user loyalty (e.g., user $u$'s visit
frequency),
$p_u(R\ne\emptyset|A)=1-\frac{\exp(\theta_u)}{\exp(\theta_u)+\sum_{i\in
A}\exp(\phi_u^\top\psi_i)}$.

\paragraph{Payoff \#2: Sales Revenue (SR).} Another important
measure of success is sales revenue or SR, which is the revenue that
a recommender system receives from the transactions (interactions)
with the users. SR is a weighted version of CTR, i.e., each click is
assigned a weight of importance. Based on CCF, SR can be formulated
via:
\begin{align}
SR&=\mathbb{E}_u[\mathbb{E}_A[\mathbb{E}_{i\in A}[c_ip_u(R=i|A)]]]\\
&=\sum_{u\in\mathcal{U}}f_u\sum_{A\in\mathcal{A}}p_u(A)\sum_{i\in
A}c_ip_u(R=i|A)\notag
\end{align}
where $c_i$ denotes the price (weight) of an item $i$.

\paragraph{Payoff \#3: Consumptions Diversity (CD).} It is widely
believed that recommender systems are the key contributor that turns
the industry from what used to be a highly concentrated
``blockbuster" \footnote{The well-known 80-20 rule or the Pareto
principle states that, of the many goods available, consumptions are
concentrated on a small subset of bestselling ones.} towards a
highly diversified long-tail (niche) market
\cite{BryHuSmi03,TanNet10}. Recent research shows that this is,
however, not entirely true --- a recommender system, if designed
improperly, could reinforce consumption concentrations
\cite{FleHos07}. In order not to turn our society to a echo chamber,
it is important to encourage consumption diversity (CD), i.e., to
ensure the consumptions of the whole population are not narrowly
concentrated. Moreover, CD is also important to online firms to help
them gain profit from long-tail market. CD can be formulated based
on the CCF framework in terms of expected choice entropy:
\begin{align}
CD
&=\mathbb{E}_u[\mathbb{E}_A[H_u(R|A)]]\\
&=-\sum_{u\in\mathcal{U}}f_u\sum_{A\in\mathcal{A}}p_u(A)\sum_{i\in
A}p_u(R=i|A)\log p_u(R=i|A)\notag
\end{align}
where $H_u(R|A)=\sum_{i\in A}p_u(R=i|A)\log p_u(R=i|A)$ is the
entropy of user $u$'s choice in the context of $A$. Note that
consumption diversity is an aggregate measure (i.e., the diversity
of the consumptions of the whole population), which is different
from the traditional individual diversity (i.e., the dissimilarity
of items recommended to an individual user).
\subsection{Implications of CCF and Future Work}
We finally remark that there are some interesting properties of the
proposed CCF model. Firstly, since the games in user-system
interactions are finite, there exists an equilibrium point (i.e., a
stable strategy). As a matter of fact, since that the reaction to a
given action is rational and that the action given $p_u(R|A)$ is
vNM-rational, it can be shown that the CCF model approximates the
\emph{perfect Nash equilibrium} \cite{LeySho08}. From a practical
point of view, it is, however, possible to optimize the recommender
systems more aggressively beyond the market equilibrium.
Particularly, the analogy of recommender system to a monopoly market
provides a number of important perspectives , e.g., the reflection
of \emph{price discrimination} in recommender system
--- how recommender system can exploit its \emph{market power} to transfer
the \emph{consumer surplus} \cite{BryHuSmi03}. Another interesting
topic is to explore the correlation and conflict of goods, and
optimize action $A$ as a bundle based on portfolio theory
\cite{Mar70,BakBry00}. We would rather leave these interesting
discussions for future research.

\section{Experiments}
\label{sec:exp}
We test the proposed CCF framework on a real-world commercial
recommender system. Because CCF is comprised of two components, it
is necessary to test each of them separately --- otherwise, it would
be difficult to tell if a change of performance is due to one
component or the other or both. Our experiments therefore consist of
two test-beds. Firstly, we compare the proposed conditional reaction
model (i.e., the Multinomial logit factor model) in our CCF
framework (referred to as CCF II) with the plain CCF preference
model proposed in our prior work \cite{YanLonSmo11} (referred to as
CCF I\footnote{Note that the essential differences between CCF I and
II are merely: (1) CCF II models null reactions and response
propensities; (2) CCF II models position bias.}) as well as
state-of-the art CF baselines in terms of their abilities in
preference estimation; to maintain a fair comparison,
recommendations are done without action optimization for CCF II,
i.e., via simple utility-based ranking. This comparison gives us an
idea on how effective our MLF model for $p_u(R|A)$ is compared with
state-of-the art preference models. Furthermore, we compare the CCF
framework (i.e., MLF + Action optimization) and the conventional
recommendation scheme (i.e., CF + utility-based ranking). This
comparison further demonstrates how the game-theoretic formulation,
particularly how action optimization, further enhance the
recommendation performance.

\subsection{Data}
We collected a large-scale set of user-system
interaction traces from a commercial News article recommender
system. In each interaction, the system offers four personalized
articles to the visiting user, and the user chooses one of them by
clicking to read that article. The recommendations are dynamically
changing over time even during the user's visit. The system
regularly logs every click event of every user visit. It also
records the articles being presented to users at a series of
discrete time points. To obtain the action set for each user-system
interaction, we therefore trace back to the closest recording time
point right before the user-click, and we use the articles presented
at that time point as the action set for the current session. We
collected such interaction traces from logged records of over one
month.  We use a random subset containing 3.6 million users, 2500
items and over 110 million interaction traces. Learning an effective
recommender on this data set is particularly challenging as the
article pool is dynamically refreshing, and each article only has a
lifetime of several hours --- it only appears once within a
particular day, is then pulled out from the pool afterward and never
appears again.

\subsection{CCF Without Action Optimization}
\label{sec:ccf-experiments}
We first evaluate CCF without action optimization (CCF II) with
comparison to the plain CCF preference model (i.e., CCF I) and the
two CF models described in Section~\ref{sec:cf}, where
recommendations are made by utility-based ranking. We consider the
following two evaluation settings, one offline and the other online.
\begin{description*}
  \item[Offline evaluation] We evaluate the learned recommender models in terms
  of the top-$k$ ranking performance on a hold-out test subset. We use
  three standard information retrieval measures as evaluation metrics, i.e.\
  average-precision at position $n$ (AP@n), average-recall at $n$ (AR@n) and
  normalized-discounted-cumulative-gain at $n$ (nDCG@n), where $n=4$, the
  default recommendation size used in the news recommender system.
\item[Online evaluation] We further conduct an online test. In
  particular.  for each incoming interaction, we use the trained
  models to predict user choice reaction, i.e., which item among the four recommended ones will
  be taken by the user. This prediction directly assesses the accuracy of the MLF
  model in user reaction modeling.
\end{description*}

\begin{table}
  \caption{Offline test: comparison of top-$k$ ranking performance.
  \label{tab:offline}}
  \medskip
\centering
\begin{tabular}{|c|c|c|c|}
\hline {Model}  &{AP@4}    &{AR@4}    &{nDCG@4}\\
\hline \multicolumn{4}{|c|}{{\textsf{30\% Training}}}\\
\hline  {CF}-$\ell_2$      & 0.245            & 0.261         & 0.255   \\
        {CF}-Logistic      & 0.246            & 0.263         & 0.257   \\
        CCF I       & 0.262            & 0.278         & 0.274   \\
        CCF II       & {\bf 0.267}      & {\bf 0.279}   & {\bf 0.278} \\
\hline \multicolumn{4}{|c|}{\textsf{{50\% Training}}}\\
\hline  {CF}-$\ell_2$      & 0.250            & 0.273         & 0.268  \\
        {CF}-Logistic      & 0.252            & 0.276         & 0.269  \\
        CCF I       & 0.266            & {\bf 0.285}         & 0.278   \\
        CCF II       & {\bf 0.269}      & { 0.284}   & {\bf 0.281} \\
\hline \multicolumn{4}{|c|}{\textsf{{70\% Training}}}\\
\hline  {CF}-$\ell_2$      & 0.253            & 0.275         & 0.271  \\
        {CF}-Logistic      & 0.253            & 0.276         & 0.274  \\
        CCF I      & 0.267            & {\bf 0.287}         & 0.280   \\
        CCF II      & {\bf 0.271}      & { 0.284}   & {\bf 0.282} \\
\hline
\end{tabular}

\caption{Online test: comparison of conditional reaction prediction
accuracy.
  \label{tab:online}}
  \medskip
\centering
\begin{tabular}{|c|c|c|c|}
\hline{Model}&{30\%train}   &{50\%train}    &{70\%train}\\
\hline{Random}&\multicolumn{3}{|c|}{0.250}\\
\hline
        {CF}-$\ell_2$      & 0.337        & 0.343         & 0.347   \\
        {CF}-Logistic      & 0.341        & 0.345         & 0.347\\
        CCF I           & 0.377            & 0.385         & { 0.391}   \\
        CCF II           & {\bf 0.383}  &{\bf 0.387}    & {\bf 0.392}   \\
\hline
\end{tabular}

\end{table}

\paragraph{Offline test results.} In this setting, we train each model on progressive
proportions of 30\%, 50\% and 70\% randomly-sampled training data
respectively, and evaluate each trained model in terms of offline
top-$k$ ranking performance. The results are reported in
Table~\ref{tab:offline}. Since the data set is fairly large the
standard deviations of all values are considerably below 0.001.
Consequently we omitted the latter from the results. As can be seen
from the table, the CCF II (i.e., MLF) model dramatically outperform
the two CF baselines in all of the three evaluation metrics.
Specifically, CCF II gains up to 9.0\% improvement over the two CF
models in terms of average precision; up to 6.9\% in terms of
average recall and up to 8.9\% in terms of nDCG. Moreover, by
modeling position bias and response propensity, CCF II also
outperforms CCF I in most (7 out of 9) of the comparisons. Note that
even compared to CCF I, the improvements achieved by CCF II are also
significant (e.g., for the system we worked on, any improvement of
the dashboard metrics especially nDCG or CTR greater than 0.1\% is a
significant breakthrough). Also note that the offline results
obtained by CCF are quite satisfactory. For example, the average
precision is up to 0.271, which means, out of the four recommended
items, on average 1.1 are truly ``relevant" (i.e.\ actually being
clicked by the user). This performance is quite promising especially
considering that most of the articles in the content pool are
transient and subject to dynamically updating.

\begin{figure}
\centering
\includegraphics[width=.75\columnwidth]{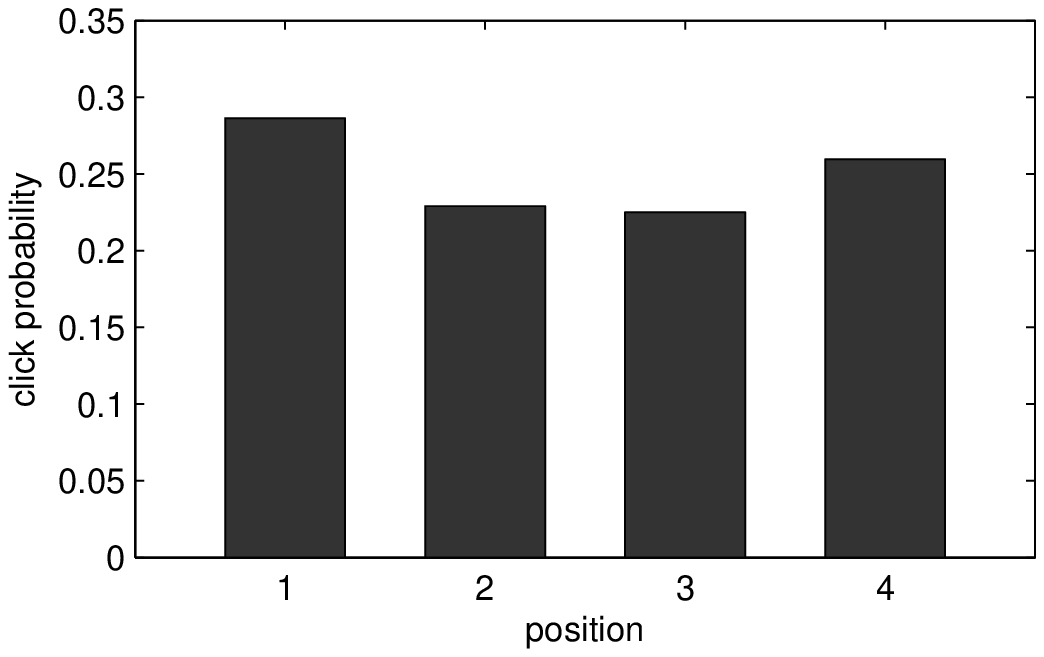}
\includegraphics[width=.75\columnwidth]{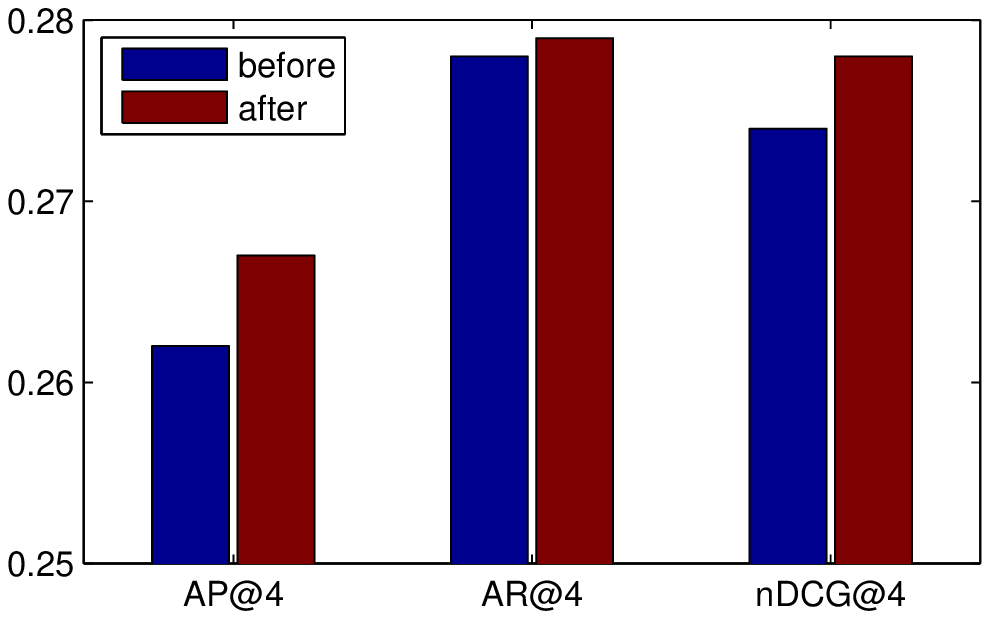}
\caption{Top: Position bias in user choice reaction; Bottom:
Comparison of recommendation performance before and after modeling
position bias. \label{fig:pb}}
\end{figure}

\paragraph{Online test results.} We further evaluate the online performance
of each compared model by assessing their predictions of user
reaction. In particular, for each of the incoming responded visits
$(u_t,A_t,i^*_t)$, we ask the question: ``among all the recommended
items $i\in A_t$, which one will most likely be clicked?" We use the
trained model to rank the items in $A$, and compare the top-ranked
item with the actual choice of the user (i.e.\ $i_t^*$). We evaluate
the results in terms of the prediction accuracy. The results are
given in Table~\ref{tab:online}. Because the size of each offer set
in the current data set is 4, a random predictor yields 0.25. As can
be seen from the table, while both the two CF models and the two CCF
models obtain significantly better predictions than the random
predictor, the two CCF models further dramatically outperform the
two CF baselines, with CCF II performs consistently the best. In
particular, CCF II improves the reaction prediction accuracy:
compared with the least square CF by 13.7\%, with the logistic CF by
12.7\% and with CCF I by 1.6\%. According to a $t$-test with
significance level 0.01, all the improvements are statistically
significant.

\paragraph{Impact of position bias.}
We observe significant position bias in the News recommender system.
As shown in Figure~\ref{fig:pb}(top), the left-most and right-most
positions (i.e., position 1 and 4) receive significantly higher
click rate than the two middle ones (i.e., position 2 and 3). In the
bottom figure, we show the recommendation performance of the CCF
(i.e., MLF) model before and after incorporating bias factors (i.e.,
$\beta$ in Eq(\ref{eq:pb})). We can see from this figure that the
performance improvements from CCF I to CCF II can be attributed
mostly to the position bias factor. Further experiments confirm that
the propensity factor only contributes a marginal improvement in
nDCG.

\begin{figure}
\centering
\includegraphics[width=.75\columnwidth]{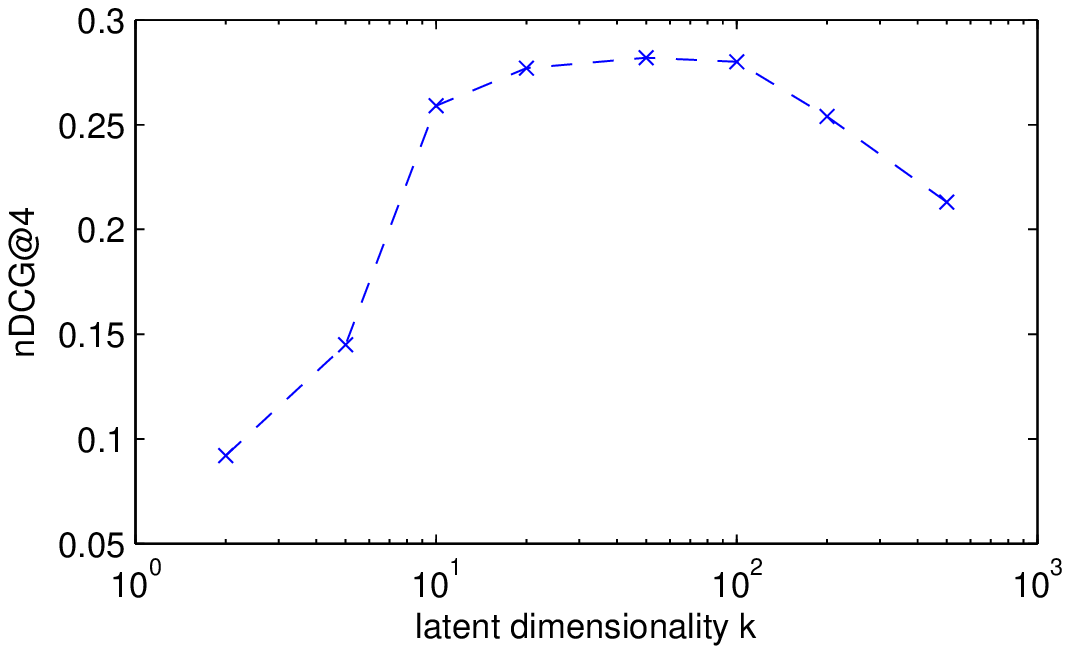}
\includegraphics[width=.75\columnwidth]{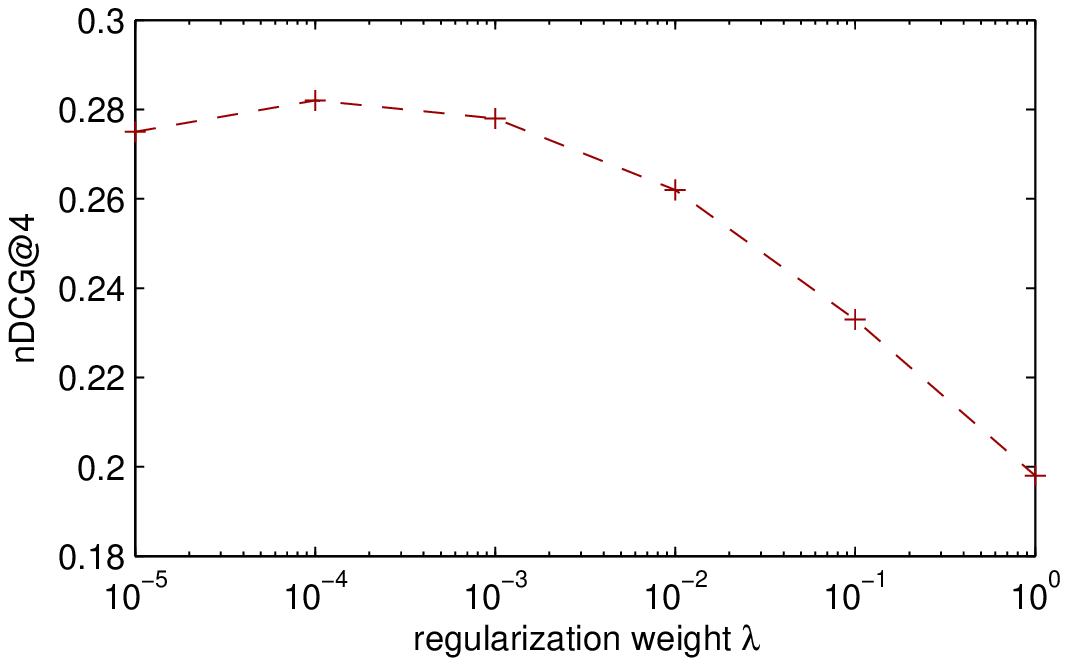}
\caption{Offline top-$k$ ranking performance (nDCG@4) as a function
of latent dimensionality $k$ (top) and regularization weight
$\lambda$ (bottom). \label{fig:para}}
\end{figure}

\paragraph{Impact of parameters.}
The performance of the MLF model is affected by the parameter
settings of the latent dimensionality, $k$, as well as the
regularization weights, $\lambda_\mathcal{I}$ and
$\lambda_\mathcal{U}$. In Figure~\ref{fig:para}\footnote{Due to
heavy computational consumptions, these results are obtained on a
relatively small subset of data.}, we illustrate how the offline
top-$k$ ranking performance changes as a function of these
parameters, where we use the same value for both
$\lambda_\mathcal{I}$ and $\lambda_\mathcal{U}$. Here we only
reported the results with nDCG@4 measure because the results show
similar tendency when other measures (including the reaction
accuracy) are used. As can be seen from the Figure, the nDCG curves
are typically in the inverted U-shape with the optimal values
achieved at the middle. In particular, for the MLF model, the
dimensionality around 50--100 and regularization weight around
0.0001 yield the best performance, which is also the default
parameter setting we used in obtaining the results reported in the
current paper.

\subsection{CCF With Action Optimization}
We now move on to evaluate the entire CCF framework (i.e., MLF +
action optimization) in terms of its ability to achieve the three
strategic goals.

\paragraph{Evaluation metrics.}
We test a recommendation model by applying it on top of the
algorithm in production and comparing the results with the
production baseline. To assess performance, we report the relative
surplus. In particular, let $m$ denote one of the three measures
(i.e., click-through rate, sales revenue and consumption diversity),
a relative surplus score is defined by:
\begin{align*}
\text{relative surplus} =
\frac{m(\text{model})-m(\text{production})}{m(\text{production})}
\end{align*}

\paragraph{Evaluation protocol.}
To illustrate how effective action optimization could be, we compare
\emph{CCF with action optimization} (CCF+AO), to \emph{CCF without
action optimization} (CCF-AO) as well as the conventional
recommendation scheme (\emph{collaborative filtering with
utility-based ranking} or CF+RK). For each model, we simulate its
relative surplus score by applying the model to the production
output. In particular, we take the top 50K users who visit our
website most frequently as test probes and trace them for one month.
For each of these user $u$ and each of the dates $d$, we maintain a
positive set $P_{u,d}$ and a negative set $N_{u,d}$ by including all
the articles that user $u$ reads on date $d$ into $P_{u,d}$ and any
other items in the content pool of date $d$ into $N_{u,d}$. We
assume user $u$ turns to take items only from $P_{u,d}$ and ignores
those in $N_{u,d}$ on date $d$. Specifically, the reaction of user
$u$ on date $d$ to any action $A$ is assumed as follows: for any
item $i\in A$, if $A\cap P_{u,d}\ne\emptyset$ and $i\in A\cap
P_{u,d}$, $u$ takes $i$ with probability $1/|A\cap N_{u,d}|$ or
otherwise ignores it; a nonresponded session occurs when
$P_{u,d}=\emptyset$. To compute sales revenue, we randomly assign to
each item a positive number as ``price", which is predefined and
never changed throughout the evaluation. Moreover, maximizing
consumption diversity alone leads to meaningless random
recommendations; to this end, we impose a hard constraint to ensure
that the decreases in CTR is no more than 0.5\%.

\begin{figure}
\centering\mbox{\subfigure{
\includegraphics[width=.9\columnwidth]{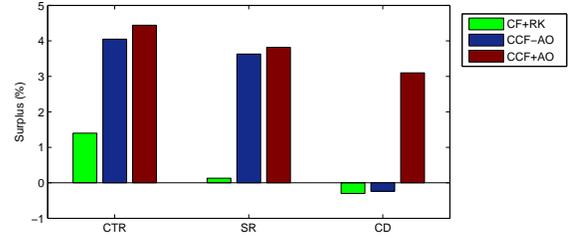}}
} \caption{Performance in achieving strategic objectives: relative
surplus compared to the production baseline in terms of CTR, SR and
CD. \label{fig:surplus}}
\end{figure}

\paragraph{Results and analysis.} The aggregate results on the 50K
probe users are depicted in Figure~\ref{fig:surplus}. Applying a
traditional recommendation scheme (CF + preference based ranking) on
top of the production baseline only yields marginal improvements in
CTR and SR. In contrast, CCF gains up to 4.5\% and 3.9\% surplus in
CTR and SR respectively; and action optimization further
significantly enhance these numbers. Interestingly, in terms of
consumption diversity, our experiment confirms the findings of
\cite{FleHos07}. For example, applying CF and CCF-AO directly
without consideration of CD inevitably leads to consumption
concentration, as shown by the negative surplus scores in
Figure~\ref{fig:surplus}. In contrast, CCF + AO is the only one
among the three models that yields positive surplus in CD. In
particular, with less than 0.5\% reduction of CTR, it gains up to
3.2\% improvement of diversity. These observations are somewhat
surprising considering that the preliminary action parametrization
we used in the experiment is a bit overly-simplistic --- it merely
contains one single parameter $\alpha$ for simple randomization
(c.f. Section~\ref{sec:actpara}). In future work, we plan to explore
more flexible forms of action parametrization such as the sequential
factorization model mentioned in \ref{sec:actpara}; we expect to
have even more promising results.

\section{Summary}
\label{sec:sum}
We presented a novel game-theoretic framework for recommendation by
viewing the user-system interactions at recommender system as
buyer-seller interactions in a monopoly economic market. Since the
decisions of the user and the buyer are interdependent, this new
perspective motivates us to optimize the action strategy of the
system by first predicting users' reaction and then adapting its
action to maximize the expected payoff. The extended CCF framework
consists two essential components: (1) a model for $p_u(R|A)$ that
integrates choice models in econometrics and latent factor model in
collaborative filtering to encode the notion of collaborative games;
and (2) a formulation for optimizing system action $A$ in terms of
expected strategic payoffs such as click-through rate, sales revenue
and consumption diversity. Experiments on a real-world commercial
recommender system have demonstrated the effectiveness and appealing
promise of the proposed framework.

\end{document}